# Multi-Task Bacterial Colony Detection and Classification Using YOLOv8 with Edge Optimization for Resource-Constrained Deployment

Belaguppa Manjunath Ashwin Desai
School of Engineering
Dayananda Sagar University
Bengaluru, India
ashwin.desai@ieee.org

Rohan Rajesh
School of Engineering
Dayananda Sagar University
Bengaluru, India
rohanrajesh0504@gmail.com

Shreyas Murthy
School of Engineering
Dayananda Sagar University
Bengaluru, India
shreyasmurthy856@gmail.com

Pronama Biswas
School of Basic and Applied Sciences
Dayananda Sagar University
Bengaluru, India
pronama-sbas@dsu.edu.in

Revathi Vaithiyanathan
School of Engineering
Dayananda Sagar University
Bengaluru, India
revathi-cse@dsu.edu.in

**Abstract—Manual counting and classification of bacterial colonies are critical yet labor-intensive tasks in microbiology, prone to human error particularly on densely populated plates. This work proposes a multi-task deep learning framework trained on the Annotated Germs for Automated Recognition (AGAR) dataset (18,000 images; 9,202 training / 3,067 testing) to automate Colony Forming Unit (CFU) enumeration and species classification. A custom multi-task CNN employing global regression served as the baseline, but demonstrated limited performance in clustered colony environments due to the absence of spatial localization. To address this, a YOLOv8 object detection architecture was adopted with high-resolution 1024×1024 inputs, enabling instance-level colony detection and label assignment. The model achieved a classification accuracy of 98.13% and a counting accuracy of 98.27% (within a 10-colony margin), demonstrating strong predictive capability. To bridge the gap between model performance and practical deployability, the trained model was optimized through unstructured and structured pruning, ONNX conversion, and reduced-precision inference (FP32, FP16, INT8). On a Raspberry Pi 4B, ONNX FP32 and FP16 variants offered the best balance between inference speed (~6.4s) and accuracy (MAE ~2.20). Unstructured pruning preserved predictive accuracy (MAE ~2.01) without runtime gains, while structured pruning resulted in significant accuracy degradation (MAE ~6.3), revealing the sensitivity of instance-level colony detection to architectural compression. These findings provide practical guidance for selecting optimization strategies in resource-constrained laboratory deployments.**

***Keywords— Bacterial Colony Enumeration, Colony Forming Units, YOLOv8, Deep Learning, Edge Deployment, Post-Training Quantization, ONNX, Raspberry Pi, Resource-Constrained Hardware***

## I. INTRODUCTION

Accurate quantification of bacterial populations through Colony Forming Unit (CFU) enumeration is a foundational step in pharmaceutical quality control, clinical diagnostics, and food safety research. In most laboratory settings, this process still relies on manual inspection under optical magnification—an approach that is slow, physically demanding, and susceptible to observer fatigue and inter-analyst variability, particularly when Petri dishes contain densely clustered or overlapping colonies [1]. The need for a reliable, automated alternative is therefore well established.

Classical image-processing approaches, including Otsu's thresholding [2] and watershed segmentation [3], have been explored for automated colony counting. However, their performance degrades significantly under real-world conditions such as non-uniform agar thickness, uneven illumination, and colony overlap [4]. Bacterial colonies also exhibit substantial variability in shape, size, color, and transparency, making reliable segmentation through fixed rules or handcrafted thresholds inherently difficult [5].

Recent advances in deep learning have enabled more robust approaches to colony detection and classification. Convolutional Neural Networks (CNNs) have demonstrated strong feature-extraction capabilities, but global regression-based CNN architectures lack explicit spatial localization, thereby limiting their effectiveness on densely populated plates [6]. Object detection frameworks such as the You Only Look Once (YOLO) family address this limitation by treating each colony as an individual instance, enabling simultaneous localization, counting, and classification [7], [8]. In particular, YOLOv8 adopts an anchor-free detection paradigm that is well suited to objects with irregular morphology and variable scale characteristics typical of bacterial colonies [9].

Despite strong performance in controlled settings, the high computational cost of such models presents a significant barrier to practical deployment, particularly in resource-constrained laboratory environments where dedicated GPU infrastructure is unavailable [10]. Post-training optimization strategies, including model pruning, quantization, and export to the Open Neural Network Exchange (ONNX) format, offer a viable pathway to bridging this gap [11].

The main contributions of this work are as follows:

- A multi-task YOLOv8-based framework trained on the 18,000-image Annotated Germs for Automated Recognition (AGAR) dataset [4] for simultaneous CFU enumeration and bacterial species classification, achieving 98.13% classification accuracy and 98.27% counting accuracy.

- A systematic evaluation of post-training optimization strategies, quantization (FP32, FP16, INT8), unstructured and structured pruning, and ONNX export, for deployment on a Raspberry Pi 4B edge device.
- A practical analysis of the accuracy-latency trade-off across optimization variants, identifying ONNX FP32/FP16 as the optimal configuration for resource-constrained deployment.

## II. METHODOLOGY

### A. Dataset and Data Splitting

The AGAR dataset was used for all experiments. The dataset comprises 18,000 images of bacterial colonies across five clinically relevant species: *Bacillus subtilis, Candida albicans, Escherichia coli, Pseudomonas aeruginosa,* and *Staphylococcus aureus*. To improve model robustness under real-world conditions, the dataset also includes samples labeled as "Contamination" and "Agar Defects". The data was partitioned into 9,202 images for training and 3,067 for testing, encompassing both high-density plates with overlapping colonies and low-density plates containing faint or minute colonies.

### B. Image Preprocessing Pipeline

Images captured under standard laboratory conditions frequently exhibit non-uniform illumination and shadow artifacts along the Petri dish rim, which can adversely affect model feature extraction. To mitigate these artifacts, a multi-stage preprocessing pipeline was developed. The Hough Circle Transform was applied to detect the curved boundary of the agar plate, from which a circular spatial mask was generated to isolate the region of interest (ROI) and suppress background pixels. Contrast Limited Adaptive Histogram Equalization (CLAHE) was then applied to correct spatially varying illumination [12]. CLAHE partitions the image into a grid of local tiles and performs histogram equalization independently within each tile, improving the visibility of faint or low-contrast colonies without amplifying noise globally. Fig. 1 shows the visual effect of this pipeline on a sample image.

### C. Baseline Model: Custom Multi-Task CNN

A custom multi-task CNN was developed as a comparative baseline to evaluate the limitations of global regression relative to instance-level detection. The architecture consists of six convolutional blocks applied to a 224×224 input, each comprising 3×3 filters, batch normalization, ReLU activations, and max pooling for progressive spatial downsampling and feature abstraction (Fig. 2). The network then branches into two task-specific heads: a scalar regression output for colony counting and a 7-way softmax layer for species classification.

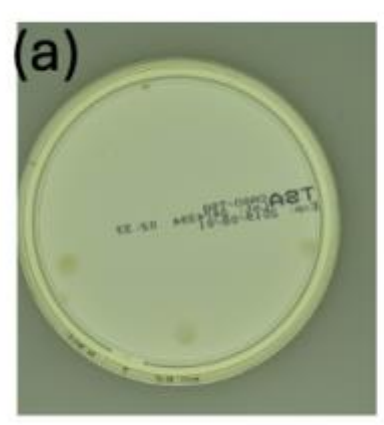

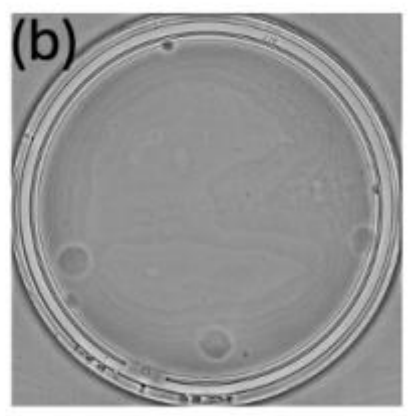

Fig. 1. Effect of the CLAHE and masking pipeline. (a) Original image with rim shadow artifacts. (b) Enhanced image after pre-processing, where faint colonies become visible.

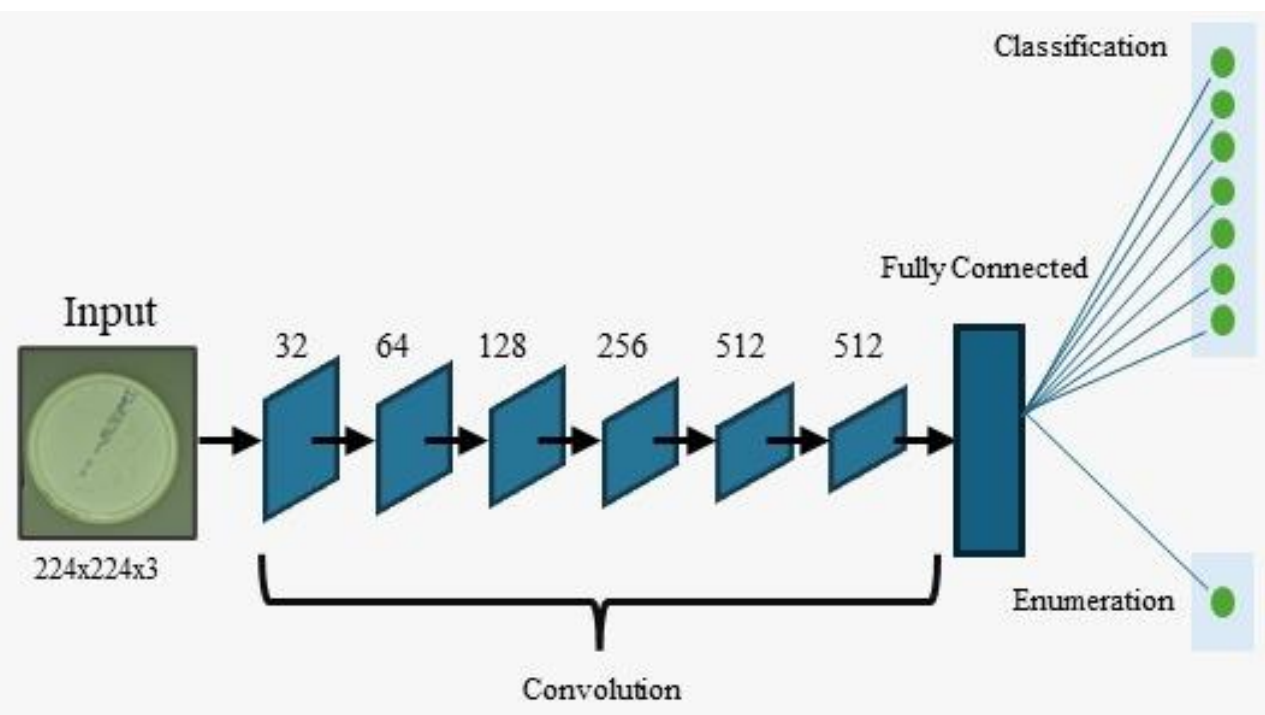

Fig. 2. CNN-based bacterial colony analysis system, illustrating the sequential stages from image input through feature extraction, pooling, flattening, and fully connected layers to final classification for 7 classes and regression output.

### D. Primary Model: YOLOv8 Object Detection

To overcome the spatial localization limitations of global regression in densely populated plates, a YOLOv8-based object detection framework was adopted. Unlike conventional detectors such as the Single Shot MultiBox Detector (SSD) [13], which rely on predefined anchor boxes, YOLOv8 employs an anchor-free paradigm that directly predicts object centers and regresses bounding box dimensions [9]. This design is particularly well-suited to bacterial colonies, which exhibit irregular morphology and large inter-class scale variability. The architecture comprises a CSPNet-derived backbone for hierarchical feature extraction and a Path Aggregation Network (PANet) neck for multi-scale feature fusion. The model was trained for 100 epochs using the AdamW optimizer with an initial learning rate of 0.001, a weight decay of 0.0005, and a batch size of 16. Mosaic augmentation was employed during training, compositing four randomly sampled images into a single training sample to improve robustness to colony overlap and varied imaging conditions (Fig. 3).

### E. Edge Hardware Constraints and Quantization

The primary objective of this work was to deploy a portable and efficient CFU analysis tool on low-cost, resource-constrained hardware. A Raspberry Pi 4B was utilized as the target edge device, running ONNX Runtime 1.24.1 for inference execution. To mitigate the computational burden of full-precision models on ARM-based CPUs, three complementary optimization strategies were implemented (Fig. 4).

*1) Quantization:* Quantization reduces model size and inference latency by replacing high-precision floating-point weights and activations with lower-precision representations (Fig. 5a). Three precision levels were evaluated: FP32 as the full-precision baseline, FP16 (16-bit half-precision floating point), and INT8 (8-bit integer quantization).

*2) Edge Device Optimization via ONNX Export*: The baseline model was exported to the Open Neural Network Exchange (ONNX) format to enable hardware-agnostic inference through ONNX Runtime (Fig. 5b). This conversion provides a standardized, optimized execution path that minimizes framework overhead by decoupling the trained model from its original training environment. Three ONNX precision variants were evaluated: ONNX FP32, which reduces inference latency relative to the native PyTorch baseline while preserving full floating-point accuracy; ONNX

FP16, which utilizes half-precision to decrease memory footprint and improve runtime efficiency; and ONNX INT8, which reduces weights and activations to 8-bit integers for maximum compression.

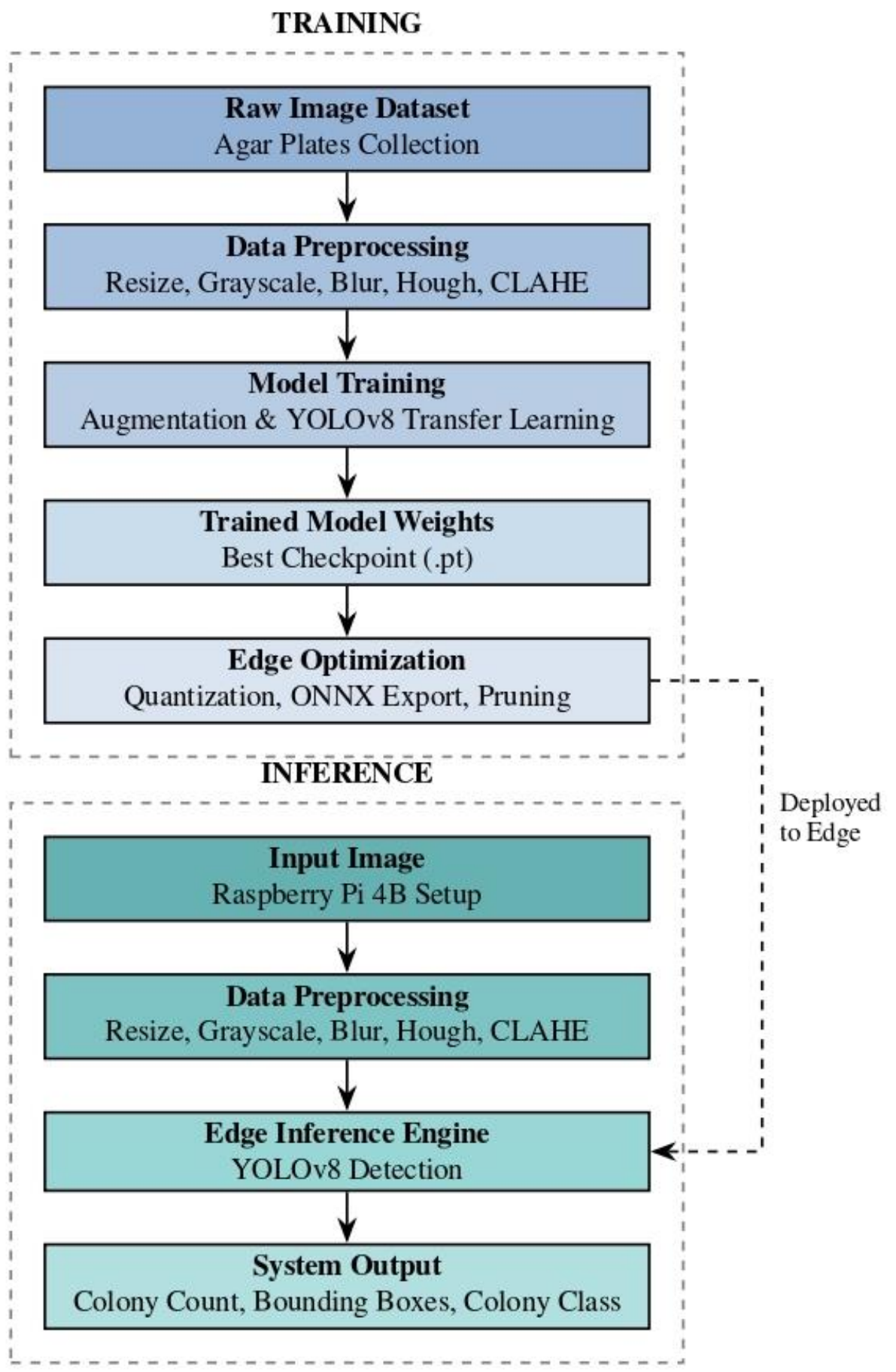

Fig. 3. End-to-end processing pipeline for the YOLOv8 bacterial colony detection system, detailing the distinct paths for model training versus edge hardware inference.

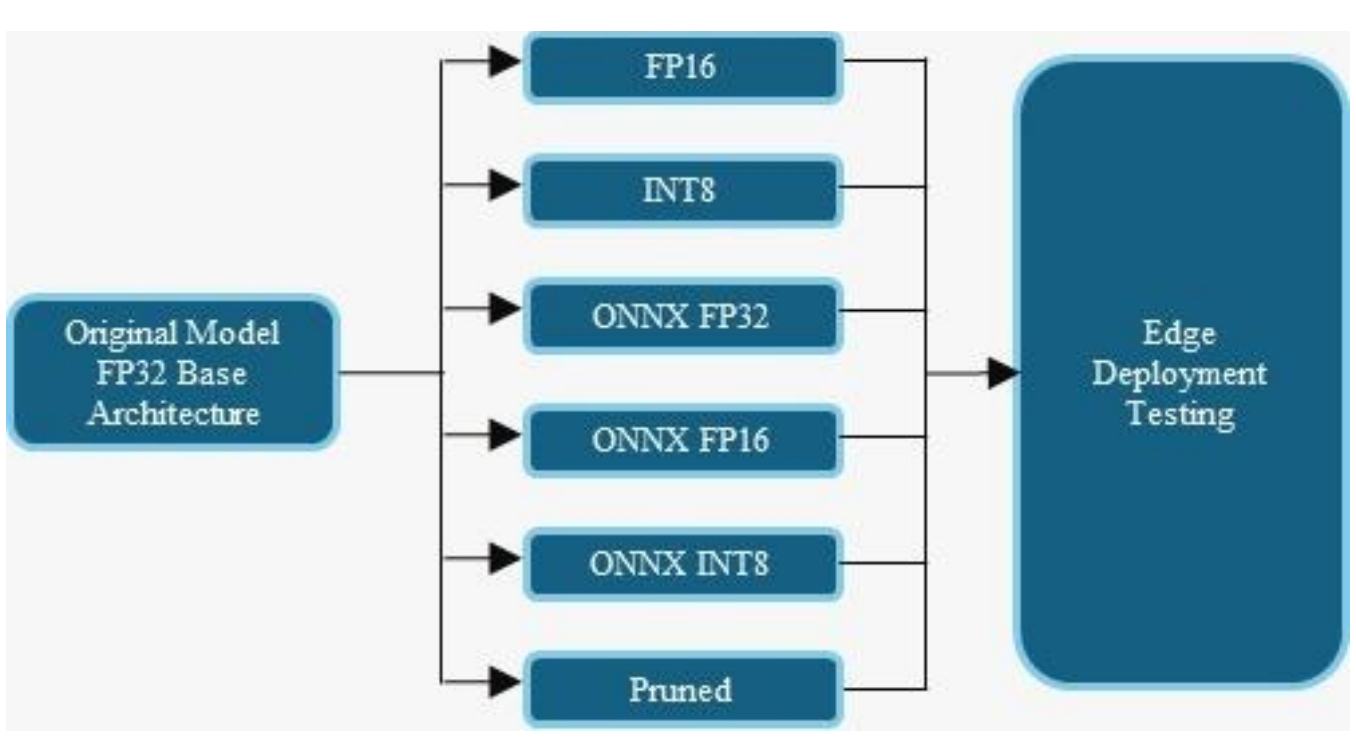

Fig. 4. Model optimization pathways for edge deployment. The baseline FP32 model was converted into FP16, INT8, ONNX-based, and pruned variants for comparative edge-deployment testing.

*3) Pruning*: Unstructured pruning was applied to reduce model complexity by zeroing out individual weights that contributed minimally to predictive capability, without altering the overall network architecture (Fig. 5c). This approach reduces the density of active weights within each layer, thereby lowering both computational cost and memory usage while preserving the original model structure.

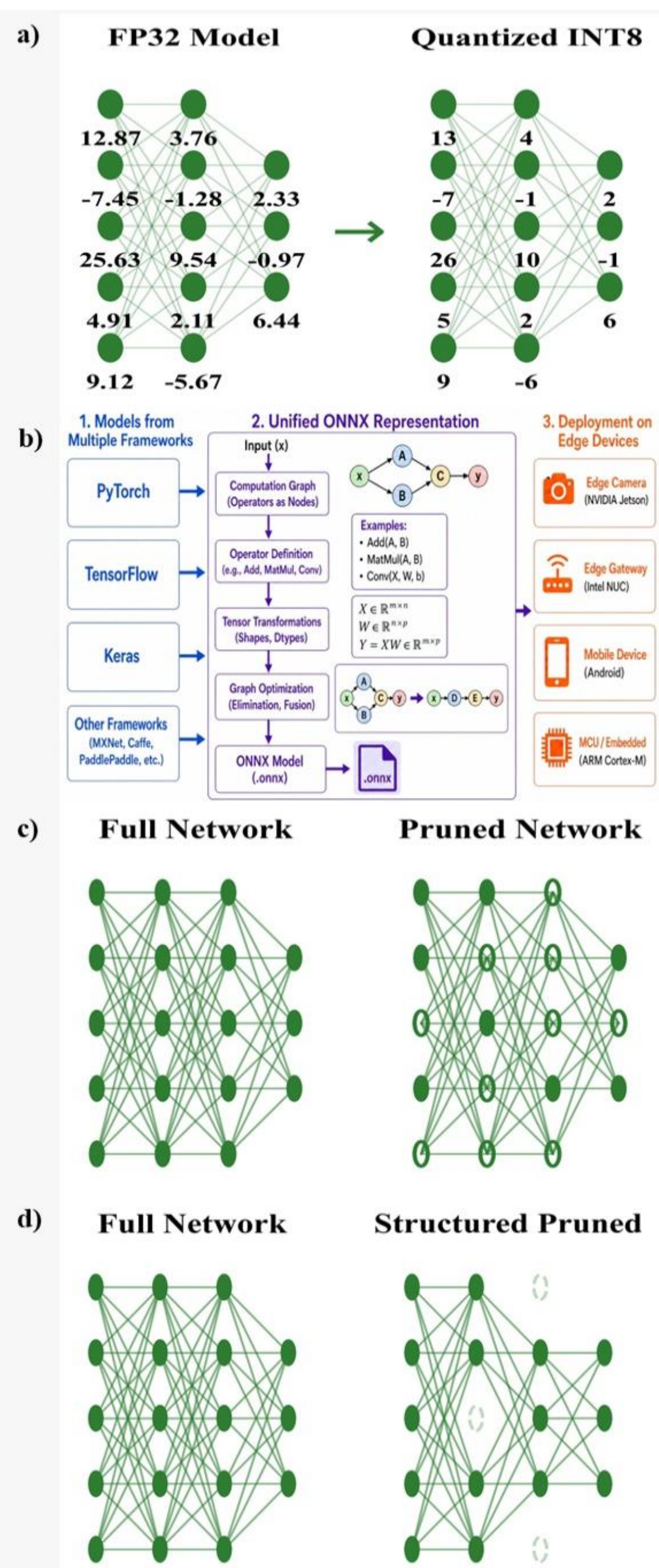

Fig. 5. Representations of various edge optimization techniques: (a) post-training quantization (FP16/INT8) for bit-depth reduction; (b) ONNX transformation for deterministic inference execution; (c) unstructured pruning; and (d) structured pruning for efficient memory utilization on resource-constrained hardware.

In addition to weight-level sparsification, structured pruning was implemented to achieve hardware-native acceleration by physically altering the network's dimensions (Fig. 5d) [14]. Unlike unstructured methods, this approach redefines the model architecture through global width-scaling to decrease filter and channel density across the backbone and head layers. This structural optimization targeted redundant filters to condense the model's complexity. To maintain functional weights without a full retraining cycle, a custom weight-migration routine was developed to transfer learned parameters from the baseline model to the scaled architecture. This was achieved by identifying and mapping only shape-compatible tensors, ensuring that the most significant features were preserved within the reduced topology. This methodology targets a direct reduction in the model's computational footprint, prioritizing a decrease in inference latency for real-time edge execution.

The final set of models evaluated, therefore, comprised the native PyTorch FP32 baseline, reduced-precision FP16 and INT8 variants, their corresponding ONNX-exported counterparts, and a pruned model. These configurations were selected to systematically examine the trade-off between inference efficiency and predictive accuracy under realistic edge deployment conditions. While quantization is known to reduce memory footprint and latency [15], the reduction in numerical precision can affect accuracy, particularly in tasks requiring the detection of small or subtle visual features.

## III. RESULTS AND DISCUSSIONS

### *A. Comparative Analysis: Regression vs Detection*

The baseline CNN model demonstrated limited performance on densely populated plates, achieving a counting accuracy of 74.1%, a classification accuracy of 64.56%, and an MAE of 3.47. Relying on global feature representations, the architecture lacked the spatial localization required to distinguish individual colonies in high-density clusters, leading to significant performance degradation under crowded conditions (Fig. 6).

Transitioning to the YOLOv8 instance-level detection framework yielded substantial performance gains. By explicitly localizing each colony as an independent object, YOLOv8 maintained high sensitivity even in overlapping environments. The model achieved a counting accuracy of 98.27% within a ±10 colony tolerance margin, a precision of 0.981, a recall of 0.966, and an MAE of 1.35 with an $R^2$ of 0.9937 against human reference counts, representing a significant improvement over the regression-based baseline.

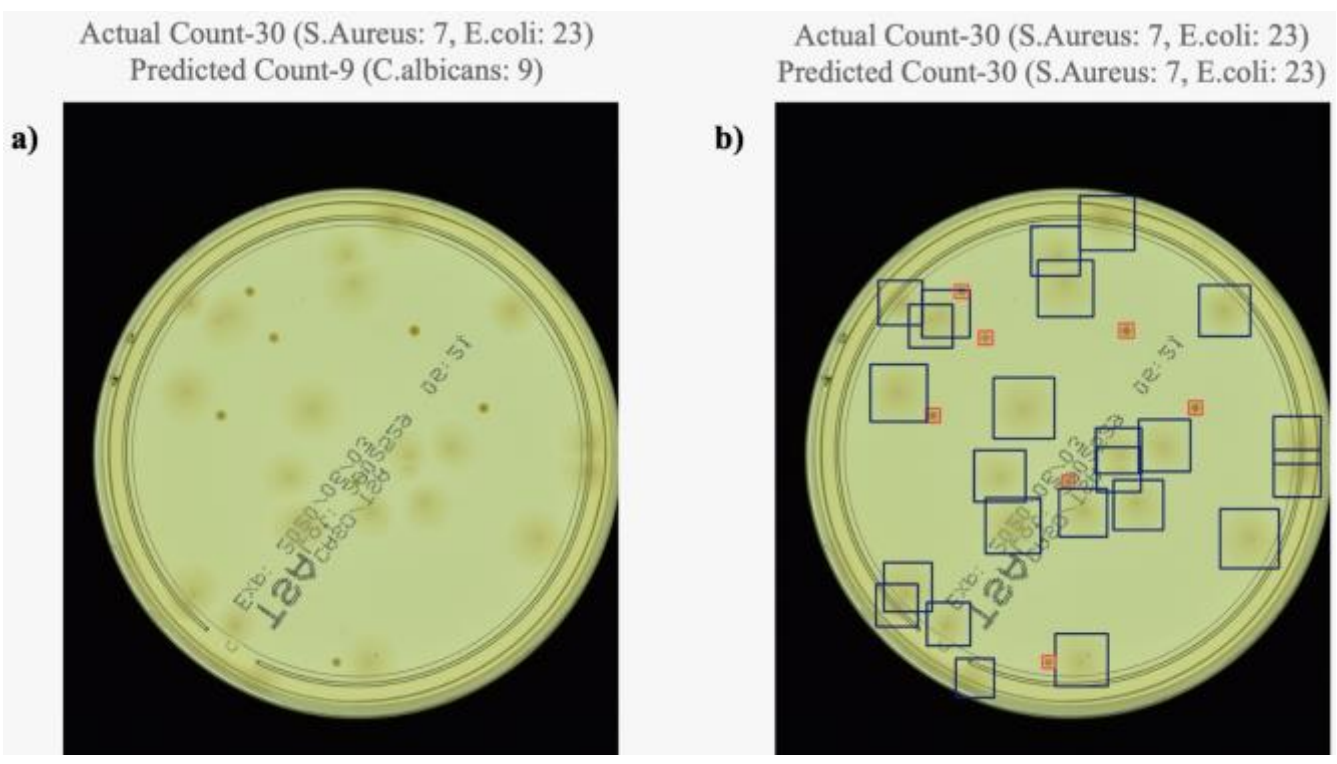

Fig. 6. Enumeration and classification prediction using (a) CNN and (b) YOLOv8.

### *B. Comparative Ablation of Model Optimization Strategies*

All compressed model variants were evaluated on a Raspberry Pi 4B for counting (Fig. 7) and classification (Fig. 8), with results summarized in Table I. A key observation across native PyTorch variants is that while quantization effectively reduced model size, it did not yield meaningful latency improvements on ARM-based CPU hardware, a limitation addressed by ONNX conversion (Fig. 9).

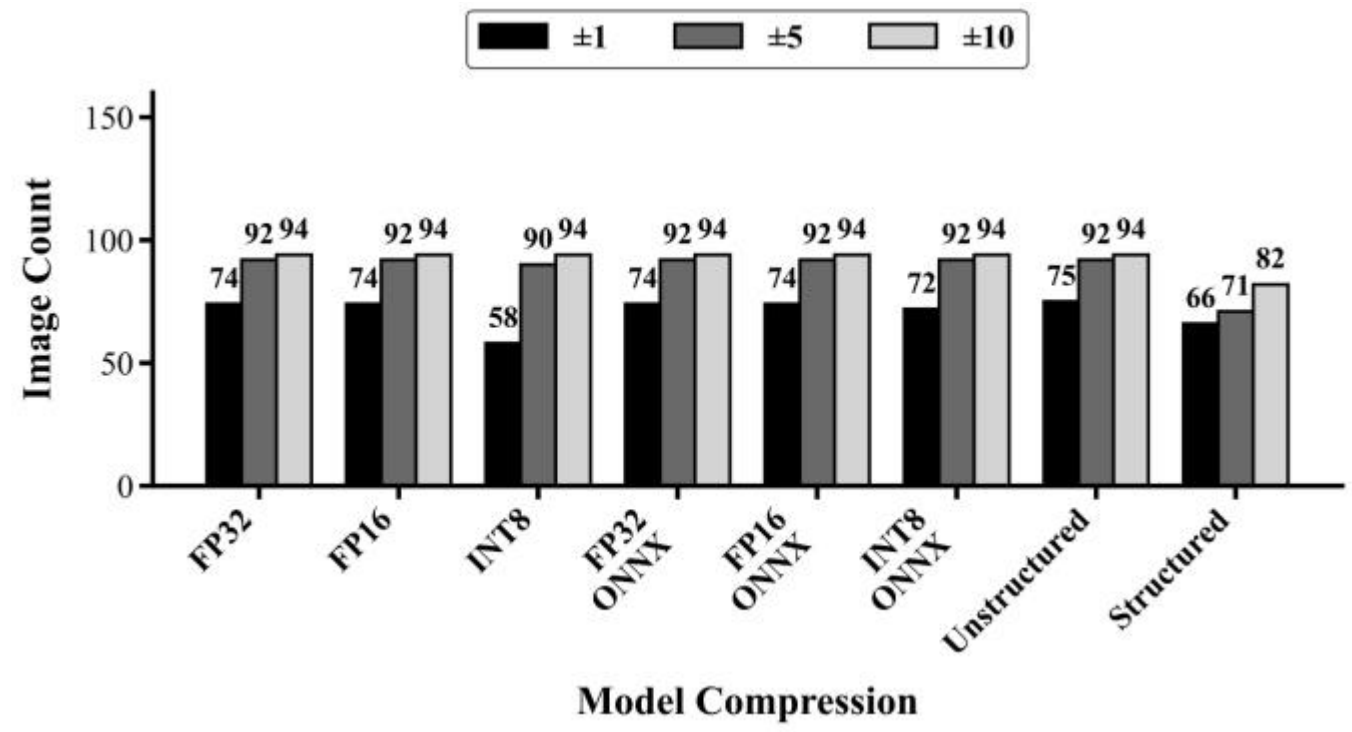

Fig. 7. Counting accuracy within specified error margins. Comparison of the number of images (out of 100) that meet counting precision targets of ±1, ±5, and ±10 colonies for each architectural and quantization optimization.

1) Quantization:

All quantized variants maintained an $R^2$ above 0.98, demonstrating strong correlation with ground-truth counts even under aggressive compression.

*FP32*: The full-precision model served as the accuracy baseline, achieving an MAE of 2.03 ± 5.54, $R^2$ of 0.9870, a classification accuracy of 98%, and a model size of 98.9 MB. However, at 7.69 ± 0.36 seconds per image, the inference latency is too high for responsive edge deployment, establishing the need for further optimization.

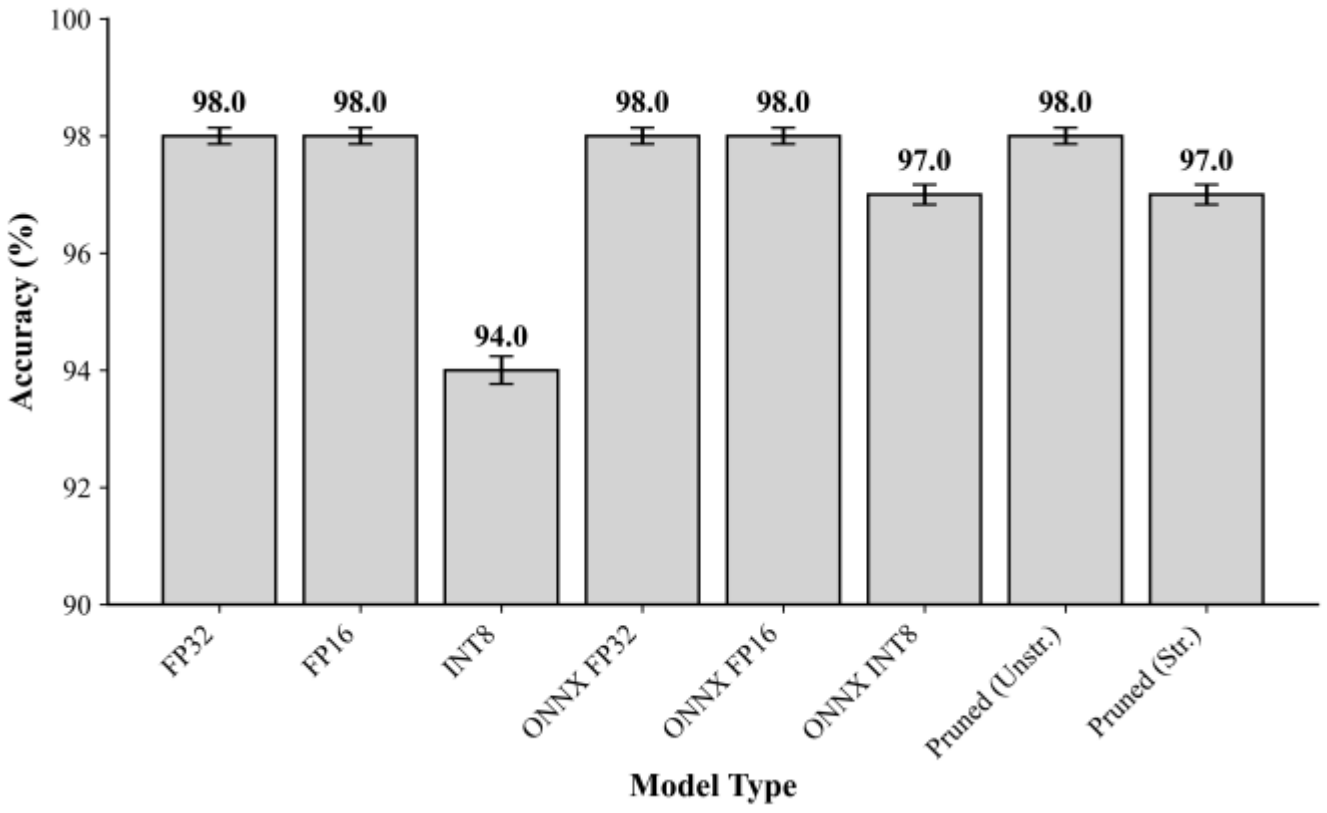

Fig. 8. Comparison of classification accuracy (%) for different model formats and optimization strategies.

*FP16:* Identical predictive performance to FP32 (MAE 2.03, $R^2$ 0.9870, classification accuracy of 98%) was preserved while model size was halved to 49.5 MB, making FP16 suitable for memory-constrained scenarios. Latency remained comparable at 7.83 ± 0.38 seconds.

*INT8:* The smallest native PyTorch footprint was achieved at 24.8 MB, with a modest accuracy trade-off (MAE 2.66, $R^2$ 0.9855, classification accuracy of 94%) and latency of 7.84 ± 0.39 seconds. Additional validation is recommended for fine-grained detection tasks, given the sensitivity of low-precision representations to subtle visual features.

2) Edge Device Optimization:

ONNX conversion provided the most significant latency improvements by delivering a standardized execution path, reducing framework overhead. All three ONNX variants outperformed their native PyTorch counterparts in inference speed.

*ONNX FP32:* Latency improved to 6.41 ± 0.07 seconds while preserving strong accuracy (MAE 2.19, $R^2$ 0.9856, classification accuracy 98%), demonstrating that ONNX conversion alone provides a meaningful deployment advantage without any precision reduction.

*ONNX FP16:* Nearly identical accuracy to ONNX FP32 (MAE 2.20, $R^2$ 0.9855, classification accuracy of 98%) at the same latency of 6.41 ± 0.05 seconds, with a reduced model size of 49.5 MB, making it the preferred choice where both memory efficiency and accuracy are required.

*ONNX INT8:* The most favorable efficiency profile overall, achieving the lowest latency of 3.71 ± 0.05 seconds and the smallest model size of 25.4 MB, with an MAE of 2.37 and $R^2$ of 0.9847, a classification accuracy of 97%, best suited for scenarios where deployment speed is the primary constraint.

TABLE I. EDGE DEPLOYMENT BENCHMARKS ON RASPBERRY PI 4B

| Model | Performance Metrics | | | | |
|---|---|---|---|---|---|
| | **MAE** | **RMSE** | **$R^2$** | **Time (s)** | **Size (MB)** |
| FP32 | 2.03±5.54 | 5.90 | 0.987 | 7.69±0.36 | 98.9 |
| FP16 | 2.03±5.54 | 5.90 | 0.987 | 7.83±0.38 | 49.5 |
| INT8 | 2.66±5.62 | 6.22 | 0.985 | 7.84±0.39 | 24.8 |
| ONNX FP32 | 2.19±5.81 | 6.21 | 0.986 | 6.41±0.07 | 98.9 |
| ONNX FP16 | 2.20±5.81 | 6.21 | 0.985 | 6.41±0.05 | 49.5 |
| ONNX INT8 | 2.37±5.94 | 6.40 | 0.985 | 3.71±0.05 | 25.4 |
| PRUNED (Unstructured) | 2.01±5.54 | 5.89 | 0.987 | 7.84±0.38 | 98.9 |
| PRUNED (Structured) | 6.31±12.36 | 22.53 | 0.928 | 7.53±0.36 | 49.0 |

3) Pruning:

Unstructured pruning achieved the lowest MAE of 2.01 among all optimized variants ($R^2$ 0.9870, a classification accuracy of 98%), closely matching the FP32 baseline in accuracy. However, since the sparsity pattern was not physically compressed into a sparse execution format, inference latency remained at 7.84 ± 0.38 seconds, and model size was unchanged at 98.9 MB. Pruning, therefore, offers accuracy preservation rather than deployment speed gains in this setting, and requires dedicated sparse inference hardware or software support to realize runtime benefits.

In contrast, structured pruning was used to physically reduce the model size by scaling the network width to 73.5% of its original capacity. This specific threshold was identified as the stability limit for the architecture; empirical testing revealed that scaling below the 73.5% mark caused the model to break. To maintain functionality at this limit, a custom weight-migration routine was developed to transfer shape-compatible parameters from the baseline model directly into the smaller architecture. This ensured that the new model retained as much learned information as possible without requiring a full retraining cycle.

This produced an MAE of 6.31 and a classification accuracy of 97%. While the accuracy was lower than that of the unstructured variant, the model maintained an $R^2$ of 0.9283, demonstrating retention of the dataset's primary predictive trends despite the diminished number of filters. These results establish a baseline for how a structurally efficient model performs when optimized strictly for architectural leanness, providing a benchmark for one-shot scaling before any further fine-tuning.

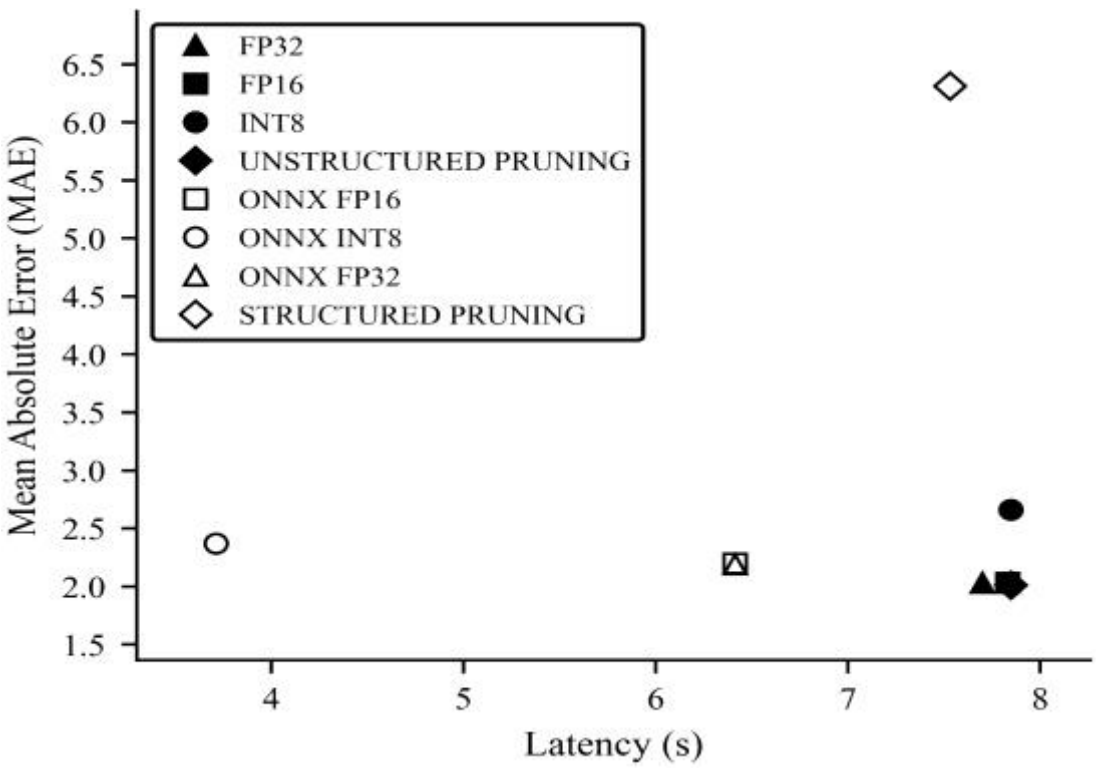

Fig. 9. Comparison of Mean Absolute Error (MAE) vs inference latency for various model quantization, pruning, and ONNX runtime configurations.

## IV. CONCLUSION

This study presented a multi-task deep learning framework for automated bacterial colony enumeration and classification, demonstrating that high accuracy and practical edge deployability can be achieved simultaneously through targeted model optimization. The YOLOv8-based detection framework trained on the 18,000-image AGAR dataset achieved a counting accuracy of 98.27% and a classification accuracy of 98.13%, substantially outperforming the global regression-based CNN baseline (74.1% counting accuracy, 64.56% classification accuracy, MAE 3.47) by virtue of instance-level spatial localization.

Systematic evaluation of post-training optimization strategies on a Raspberry Pi 4B revealed that ONNX conversion is the single most impactful optimization for reducing inference latency on ARM-based CPU hardware. ONNX FP32 and FP16 variants offered the best balance between accuracy (MAE ~2.20, $R^2$ ~0.9856, classification accuracy of 98%) and inference speed (~6.4 seconds), while ONNX INT8 achieved the lowest latency of 3.71 seconds at a modest accuracy trade-off (MAE 2.37), making it suitable for speed-critical deployments. Native PyTorch quantization reduced model size effectively but did not yield latency improvements, and unstructured pruning preserved accuracy (MAE 2.01) without runtime gains, highlighting that hardware-aware optimization strategies are essential for effective edge deployment. In contrast, structured pruning caused a substantial accuracy degradation (MAE 6.31, $R^2$ 0.9283, classification accuracy 97%) without delivering proportional latency benefits, indicating that naive structural sparsification may not be advantageous on general-purpose ARM CPUs without dedicated acceleration support.

These findings demonstrate that the proposed framework offers a portable, cost-effective, and dependable solution for automated CFU analysis, with the choice of optimization variant guided by the deployment context's tolerance for latency versus accuracy trade-offs. Future work will focus on quantization-aware training to recover accuracy lost under aggressive compression, structured pruning compatible with sparse inference acceleration, and robustness evaluation across varied plate geometries, colony densities, and imaging conditions.

## Acknowledgment

The authors thank the creators of the AGAR dataset for making the annotated plate images publicly available, without which this study would not have been feasible. Experiments were conducted using the Ultralytics YOLOv8 framework.

## Declaration of Generative AI Use

During the preparation of this work, the authors used an AI-assisted language model (Claude, Anthropic) to assist with citation formatting and manuscript proofreading. After using this tool, the authors reviewed and edited the content as necessary and take full responsibility for the content of this publication.